\documentclass[conference]{IEEEtran}

\usepackage{subfigure}
\usepackage{amsmath}
\usepackage{amsfonts}
\usepackage{amssymb}
\usepackage{epsfig}

\usepackage{algorithm}
\usepackage{algorithmic}

\usepackage{times}
\usepackage{amsmath}
\usepackage{graphicx}

\def\comment#1{}

\newcommand{\Prob}{\ensuremath{\mathbf{P}}}

\newcommand{\numSamp}{T}
\newcommand{\numK}{K}

\newcommand{\psme}{\eta}

\newcommand{\numB}{S}
\newcommand{\rvMisM}{X}
\newcommand{\valMisM}{x}

\newcommand{\rvErrB}{Y}
\newcommand{\valErrB}{y}

\newcommand{\rvMaxE}{Z}
\newcommand{\valMaxE}{z}
\newcommand{\numByte}{U}

\newcommand{\node}{v}
\newcommand{\pmtr}{M}
\newcommand{\load}{L}
\newcommand{\lera}{e}
\newcommand{\lerb}{q}
\newcommand{\path}{P}

\newcommand{\ovhr}{H}
\newcommand{\rcvb}{d}

\newcommand{\djset}{\pi}
\newcommand{\djcand}{w}

\def\ps@headings{%
\def\@oddhead{\mbox{}\scriptsize\rightmark \hfil \thepage}%
\def\@evenhead{\scriptsize\thepage \hfil \leftmark\mbox{}}%
\def\@oddfoot{}%
\def\@evenfoot{}}
\makeatother
\begin{document}
\title{Employing Coded Relay in Multi-hop Wireless Networks
\author{
Zhenghao Zhang, Wei Hu, and Jin Xie\\
{\normalsize Computer Science Department}\\
{\normalsize Florida State University Tallahassee, FL 32306, USA}
}
}
\maketitle

\markboth{}{}


\pagestyle{plain}
\setcounter{page}{1}

\begin{abstract}

In this paper, we study Coded relay (Crelay) in multi-hop wireless networks. Crelay exploits both partial packets and overhearing capabilities of the wireless nodes, and uses Forward Error Correction code in packet forwarding. When a node overhears a partial packet from an upstream node, it informs the upstream node about the number of parity bytes needed to correct the errors, such that the upstream node need only send a small amount of parity bytes instead of the complete packet, hence improving the network efficiency. Our main contributions include the following. First, we propose an efficient network protocol that can exploit partial packets and overhearing. Second, we study the routing problem in networks with Crelay and propose a greedy algorithm for finding the paths. Third, we propose an error ratio estimator, called AMPS, that can estimate the number of byte errors in a received frame with good accuracy at a low overhead of only 8 bytes per frame, where the estimator is needed for a node to find the number of needed parity bytes. Fourth, we implement the proposed protocol and algorithm within the Click modular router, and our experiments show that Crelay can significantly improve the performance of wireless networks.

\end{abstract}

\section{Introduction}

Wireless multi-hop networks, e.g., wireless mesh networks, wireless sensor networks, have attracted much interests in recent years. In this paper, we propose Coded Relay, abbreviated as {\em Crelay} in the following, which exploits two fundamental properties of transmissions over the wireless medium, namely the existence of partial packets and the overhearing capability. That is, partial packets are often received by wireless nodes that are not completely correct but still contain a significant amount of correct information. Also, because the medium is shared, when a node transmits a packet to second node, a third node may overhear this packet. 

The core idea of Crelay is simple and can be explained as follows. Basically, nodes relay {\em coded} messages to the next hop node depending on the amount of information that has already been overheard, where a coded message is constructed according to an Forward Error Correction (FEC) code. As a simple example, suppose a path is $\node_A \rightarrow \node_B \rightarrow \node_C$, while $\node_A$ wishes to send a packet $\Prob$ to $\node_C$. $\node_A$ first transmits $\Prob$, after which $\node_B$ gets $\Prob$ while $\node_C$ gets a partial packet with some errors. $\node_C$ estimates the number of errors using an error ratio estimator, and asks $\node_B$ to send just enough number of parity bytes correct these errors, instead of sending the entire packet. Thus, fewer bytes are transmitted and a better efficiency is achieved. By sending FEC-coded messages, Crelay opens up new possibilities for packet forwarding in multi-hop wireless networks.

Although the idea is simple, the design of Crelay faces the following challenges. First, a protocol should be designed for control message exchange that facilitates packet forwarding at low overhead and low delay. The control message should allow an upstream node to be informed about the receiving status of a packet to determine whether the packet should be transmitted and if so, the number of parity bytes needed. Note that a packet transmission may reach a far node on the path due to a lucky overhearing, and the bypassed nodes should be exempted from the duty of forwarding because their transmissions are pointless at this moment. Also, the upstream node should be aware of the number of errors in a received packet such that it can send just enough number of parity bytes to correct the errors. Avoiding pointless transmission is the classic challenge facing all opportunistic routing protocols. Existing approaches include structured forwarder coordination \cite{ExOR} which may discourage spatial reuse \cite{MORE}, or randomized network coding \cite{MORE,MIXIT} which cleverly eliminates the need of feedback but cannot be used in Crelay because Crelay needs the feedback to determine the number of parity bytes. To this end, we give a novel, simple solution, based on two key observations: (1) {\em packets usually experience queuing delays at the nodes}, specially under high load when throughput should be optimized; (2) {\em the lucky overhearing usually bypasses a small number of nodes on the path} such that it is possible to propagate the status of an overheard packet to the upstream in a timely manner. Therefore, with a good feedback mechanism, a node can often obtain the receiving status of a packet from its downstream nodes before starting to serve this packet, because it has to serve other packets first. We design an efficient feedback mechanism for Crelay which scavenges all overheard useful information and adopts two tricks we call the  {\em ACK triggered record} and {\em ACK propagation}. The overhead is low because all feedback information is piggybacked with data packets whenever possible, not necessarily the packets belonging to the same flow.

Second, an algorithm is needed to calculate the best path in Crelay. The routing problem is interesting because a sub-path of an optimal path may not be optimal, due to partial packets and overhearing. We study the routing problem and propose a practical heuristic algorithm for finding the paths. 

Third, an estimator is needed to find the number of errors in a received packet, because the receiving node actually does not know the number of errors. We propose an error ratio estimator, referred to as {\em AMPS}, which is based on the optimal maximum a posteriori (MAP) estimation. AMPS adds only 8 bytes per-frame as overhead and computes the estimate with a constant time table lookup. Our simulations also show that for per-packet level estimation, AMPS is more accurate than EEC \cite{EEC}, a recently proposed error ratio estimator, at much less overhead. Our experiments show that AMPS can achieve good accuracy, e.g., for more than 58\% of the time, its estimation error is no more than 3 bytes among typically 150 transmitted bytes. 

We implement Crelay within the Click modular router \cite{Click}, and test its performance in an 11-node testbed. The results show that Crelay achieves a significantly better performance than the traditional single path routing scheme as well as More \cite{MORE} which the state-of-art opportunistic routing protocol without physical layer hint. For example, the average throughput gain of Crelay  over  More is 36\% in our experiments.



The rest of the paper is organized as follows. 
Section \ref{relworks} discusses the related works.
Section \ref{protocol} describes the Crelay protocol. 
Section \ref{Algorithm} discusses the problem of finding paths. 
Section \ref{ErrorEst} describes the AMPS error ratio estimator. 
Section \ref{Evaluation} gives the experimental results.
Section \ref{Conclusions} concludes the paper.

\section{Related Work}
\label{relworks}


Partial packets and overhearing opportunities have been realized and studied extensively in recent years, see, for example, \cite{SIGCOMM07PPR,ZipTx,ExOR,MORE,MIXIT}. Compared to other opportunistic routing protocols, Crelay operates more like a traditional routing protocol, in that (1) Crelay forwards packets in a per-packet basis while ExOR \cite{ExOR}, More \cite{MORE}, and MIXIT \cite{MIXIT} require a batch of packets to be assembled before the transmission, (2) Crelay maintains per-neighbor buffer while ExOR, More, and MIXIT maintain per-flow buffer hence have higher protocol complexity. In addition, network-coding-based solutions such as More and MIXIT need to solve linear equations to recover every packet, which poses high requirements on the computational capabilities as well as power capabilities. Another core difference between Crelay and MIXIT is that Crelay does not rely on physical layer hints to handle partial packets, hence can be used in a wider range of application scenarios because physical layer hints are not always available \cite{ZipTx}.

Crelay uses FEC code in a network, however it is different from network coding. Wireless network coding combines multiple packets into a coded packet, usually through a linear transformation, and broadcast the coded packet \cite{firstnwcoding,MORE,MIXIT,Sigcomm07,Sigcomm06}. Crelay does not mix multiple packets and does not incur the associated computational cost. 

Introducing relaying nodes has been proposed for cellular phone networks. For example, LTE base stations may employ relaying devices in a cell which can significantly improve the throughput \cite{LTErelay1}. The idea of Crelay was actually originated from {\em corporative relaying}, where itself is still an active research area in the signal processing community \cite{relaycode01,relaycode02}. Existing research on corporative relaying considers forwarding messages in a two-hop scenario; however, Crelay focuses on multi-hop networks, where the routing problem is much more challenging.


Recently, Error Estimation Coding (EEC) \cite{EEC} was proposed for estimating the error ratios. EEC focuses on the average error ratio of a link, while APMS focuses on the individual error ratio of the packets. EEC, with non-trivial probability, may output 0 as the estimated number of errors for partial packets, which is tolerable for link-level estimation after taking the average; however, it cannot be applied directly to Crelay which needs accurate, per-packet estimation. We make a more detailed comparison between EEC and AMPS in Section \ref{compEEC}, where we show that for per-packet estimation, AMPS outperforms EEC with lower overhead.

\section{The Crelay Protocol} 
\label{protocol}

We describe the Crelay protocol in this section. 

\subsection{Preliminaries}

Crelay is basically a link state protocol where nodes learn and propagate the quality of the links. A link is measured by two values, namely the {\em erasure ratio} and the {\em error ratio}, where the former is the fraction of frames that are unaware of by the receiver and the latter is the fraction of the erroneous bytes in a received frame. A {\em path} is an ordered list of nodes that will participate in relaying the packet. All nodes appearing earlier than a node in the list are the {\em upstream nodes}, and all nodes appear later the {\em downstream nodes}. Nodes determine the packet forwarding paths based on the link states, and every node should find the same path for a particular pair of source and destination. When received a packet from the upper layer, the source node inject the packet into the network whenever it gets the opportunity and nodes in the forwarding path will collaborate in relaying it, bypassing some nodes sometimes, until it reaches the destination. A node maintains a queue for each neighbor, and buffers a packet in the queue for neighbor $\node_A$ if the next hop of the packet is $\node_A$. Packets in the a queue are first-come-first-served; different queues are served according to round-robin. 

Crelay can work with any error correction code. The current implementation uses the Reed-Solomon (RS) code, because of its strong error correction capability and the availability of software implementations \cite{khRS}.  Using the RS code, if there are $e$ erroneous bytes in the received data belonging to a codeword, with {\em any} additional $2e$ bytes in the codeword, all errors can be corrected. A node does not have to receive all bytes in the codeword; the part that are not transmitted can be treated as {\em erasures} \cite{bookWicker}.

Errors in the packets may occur in bursts, which is not desirable because it may result in one codeword handling too many errors, thus exceeding the decoding capability, while others handling too few. To cope with this, Crelay adopts {\em interleaving}. Basically, before transmission, bytes in the data field of a frame are mapped to random locations according to a random permutation, and at the receiver, the reverse of the random permutation is applied before handing the packet for processing. The effect of this is that errors are relocated to random locations such that they are spread evenly in the packet.


\subsection{Block, Codeword, Segment, and Record}

The four main concepts in Crelay with regarding to packet forwarding are {\em block}, {\em codeword}, {\em segment}, and {\em record}:
\begin{itemize}
\item Block: a data packet from the upper layer is divided into a number of blocks of equal size, padded if necessary.
\item Codeword: a block is encoded according to the RS code into a codeword, which is basically the data bytes followed by the generated parity bytes.
\item Segment: a continuous segment of bytes in a codeword, represented by an interval of integers.
\item Record: the received segment(s) in a codeword. The segment(s) may be scattered in the codeword. They usually do not overlap; in case of overlap, for the overlapping part, the one with less errors is used.
\end{itemize}
In our current implementation, each block is 150 data bytes, and each codeword is 255 bytes according to the (255, 150) RS code. When transmitting a packet, for simplicity, a node transmits the same segment in all codewords. We say a record is {\em decodable} if the original data block can be recovered from it. If a block has been recovered from the record, we also say it is decoded. If all blocks in a packet are decoded, we say the packet is decoded.

\subsection{Basic Operations}

\subsubsection{Receiver} A node always monitors the channel. When receives a data packet, if it is on the forwarding path of this packet, found based on the source and destination ID, it checks whether it is on the downstream of the packet sender. If yes, it adds the newly received segments to its records of this packet. It then runs AMPS to estimate the number of errors. As the number of errors in the transmitted segments may vary, it estimates the maximum number of errors in a segment among all segments, and uses it as the estimate for all segments.  It then estimates whether the records are decodable. If no, when gets access to the medium, it announces the receiving status of the packet, which includes the start location, end location, and estimated number of errors of each segment that has been overheard. After receives another segment, a node may attempt to decode. If the decoding is successful, it sends an ACK; otherwise it announces the new receiving status and waits for the next segment until all records are decoded or a timeout. It could happen that the decoding fails only at certain records; in this case the node announces a mask of the decoded records, and the upstream node will transmit segments only for the undecoded records. 

\subsubsection{Sender} The sender, when transmitting for a packet, chooses a minimum size segment such that the next hop node should be able to decode the records. The receiving status of nodes further downstream are not considered because they may be better served by nodes closer to them. Basically, it runs a linear search at selected locations and estimates the number of bytes needed if the segment starts at this location, and picks the one that needs the minimum. The list of locations include the start and end of each segment that  has been overheard, plus the first byte of the codeword.


\subsection{A Simplified Example}

\begin{figure}
\begin{center}
\includegraphics[width=2.5in]{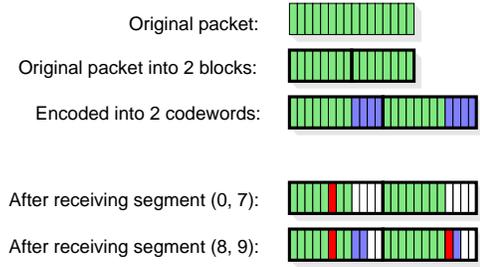}\\ 
\end{center}
\vspace{-0.15in}
\caption{An example of Crelay. Green: data bytes. Blue: parity bytes. Red: corrupted bytes. }
\label{fig:crexample}
\vspace{0in}
\end{figure}

Fig.~\ref{fig:crexample} is a simple example for the illustration of the concepts. The data packet from the upper layer is assumed to be 16 bytes, and is organized into 2 blocks. Each block is encoded into a codeword with 4 parity bytes. The first transmission is segment (0,7), i.e., the data bytes, for both blocks. The channel corrupted 1 byte in the transmitted segment for block 0. The receiving node estimates the maximum number of errors among the two transmitted segments to be 1, and announces (0,7,1) as the receiving status. The sending node, seeing that there is 1 error, transmits segment (8,9), because the 2 parity bytes should correct the error. The channel actually corrupts 1 byte in the transmitted segment for block 1. The receiving node estimates that there is no error in the received segments, which is an incorrect estimation, and believes that it has two segments in its record: (0,7,1) and (8,9,0). As 2 parity bytes should correct 1 error, it attempts to decode the records and luckily decoded both.  

\begin{figure}
\begin{center}
\includegraphics[width=3in]{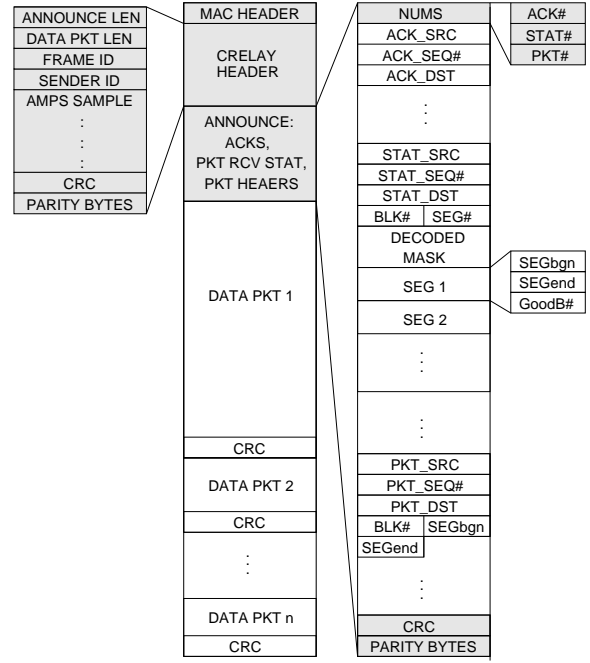}\\ 
\end{center}
\vspace{-0.15in}
\caption{Crelay frame format. The shaded fields are mandatory, others are optional.}
\label{fig:crformat}
\vspace{-0.2in}
\end{figure}

\subsection{Frame Format} 

After the MAC header, a Crelay frame consists of three main sections, the header, the announcements, and the data packets, as shown in Fig.~\ref{fig:crformat}. The header contains information such as the frame sequence number, the sender's ID, AMPS samples, etc, and is protected by an FEC code. The announcements section contains ACKs, packet receiving status, and headers of the data packets in this frame, also protected by an FEC code. The ACK contains simply the source and destination ID and the packet sequence number. The packet receiving status contains the source and destination ID, the packet sequence number, the number of blocks in the packet, the mask of decoded blocks, and the list of received segments. The data packet header contains the source and destination ID, the packet sequence number, the number of blocks, and the transmitted segment. The data packets section contains the data. A frame may have multiple data packets, because one may be a fresh packet and the other may be the parity bytes for another packet. 

The overhead of the Crelay protocol is mainly the Crelay header and the announcement section. The Crelay header is fixed 28 bytes. The length of the announcement section may vary depending on the number of ACKs, packet receiving status, and data packets, and is usually no more than 60 bytes. Our experiments show that Crelay is able to achieve improved performance over other protocols at this overhead.

\subsection{Addressing the Protocol Design Challenges} 

We now discuss how Crelay meets the main design challenges. 

\subsubsection{ACK Triggered Record}

Crelay should first ensure that a node sends the correct amount of parity bytes to the next hop node. Denote a node as $v_A$ and its next hop as $v_B$. This requires $v_B$ to get a good estimate of the number of errors, addressed by AMPS to be discussed later; also, $v_A$ should wait for the receiving status of $v_B$ before sending the packet. Crelay therefore classifies packets into three states, S0, S1, and S2: (1) in state S0, some information has been overheard about this packet, but it still not decodable; (2) in state S1, the packet has been successfully decoded, but the receiving status of the packet at the next hop is not known; (3) in state S2, the packet has been successfully decoded, and the receiving status of the next hop is known. Only packets in state S2 can be transmitted. 

If $v_B$ overhears the packet, $v_B$ should be able to announce the receiving status when it gets access to the medium. The challenge is that $v_B$ may not have overheard it hence never announces the receiving status, and $v_A$ may hold the packet in state S1 forever. Crelay solves this problem with a simple trick called ``ACK triggered record.'' That is, it let nodes create an empty record of a packet once overheard an ACK for this packet, even when no data is overheard. The empty record will prompt $v_B$ to announce an empty receiving status, which will allow $v_A$ to promote the packet into state S2. The rationale behinds this is that $v_A$ will send an ACK once it decoded the packet, and most likely, this ACK can be overheard by $v_B$ because they are neighbors on the path who should share a relatively good link. As nodes should follow a fair MAC protocol to access the channel, after $v_A$ sends the ACK, $v_B$ is likely to get the channel and be able to announce the receiving status. Thus this solution does not increase much of the delay. In the case that $v_B$ did not overhear the data packet and the ACK, Crelay relies on timeout and allows a packet to be promoted to state S2 if it has been in state S1 for longer than a threshold.


\subsubsection{ACK Propagation}

Crelay should also ensure that the ``good forwarders'' forward the packet to reduce pointless transmissions. That is, suppose a path is $v_0 \rightarrow v_1 \rightarrow ... v_{n-1} \rightarrow v_n$. After $v_i$ gets the packet due to a lucky overhearing, $v_j$ should not transmit if $j < i$. To achieve this, Crelay adopts a simple strategy based on ``ACK propagation.'' Basically, an ACK will be propagated from the downstream to the upstream whenever needed, helping the upstream nodes to remove packets that no longer need to be transmitted. 

To be more specific, in Crelay, a node sends ACK in three cases:  (1) when it decoded the packet, (2) when it removed the packet recently because received an ACK from a downstream node but heard an upstream node sending it again, and (3) when it overheard an ACK or the data packet from a downstream node while it has not decoded the packet itself. Any node who gets an ACK for a packet from a downstream node will delete the packet. Case (1) is obvious and Case (2) is because if the upstream node is still sending the packet, it did not get an ACK from any of its downstream nodes and the downstream node should send ACK again. Case (3) guarantees that if a node sends an ACK or sends the packet itself, both can be regarded as a valid ACK, this ACK will be propagated and eventually known to the upstream nodes that are waiting for ACKs. The propagation could take time, and it might be a concern that a node may make an unnecessary transmission  before the ACK is propagated to it. However, as discussed earlier, the packet queuing delay usually prevents this from happening. Also, in most cases, this lucky overhearing bypasses a small number of links so the propagation delay is usually small.

\section{Routing With Crelay} 
\label{Algorithm}

\begin{figure}
\begin{center}
\includegraphics[width=1.5in]{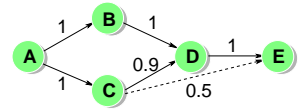}\\ 
\end{center}
\vspace{-0.15in}
\caption{Routing example with Crelay.}
\label{fig:routex}
\vspace{-0.15in}
\end{figure}

In this section, we discuss the routing problem in Crelay. Routing in Crelay is interesting because a sub-path of an optimal path may no longer be optimal. For example, consider the simple network shown in Fig.~\ref{fig:routex}. The number besides a link is the {\em receiving ratio}, defined as the number of bytes that can be decoded if the sender sends one byte. It represents the quality of the link and is determined by the error ratio and the FEC code adopted. The best path from $A$ to $D$ is clearly $A \rightarrow B \rightarrow D$, but the best path from $A$ to $E$ is $A \rightarrow C \rightarrow D \rightarrow E$, due to the overhearing link from $C$ to $E$.

\subsection{Path Metric} 
We use the {\em Expected Transmission Byte (ETB)} as the metric of a path, which is the expected of bytes needed to be sent in total such that the destination can receive one byte of data. It can be calculated based on the erasure ratio and error ratio of the links on the path. 

To be more specific, denote a path as $\path = \node_0 \rightarrow \node_1, ... ,\rightarrow \node_n$. Let $\load[\node_i]$ denote the expected load of $\node_i$, defined as the expected number of bytes $\node_i$ should send. The metric of path $\path$ is clearly 
$
\pmtr[P] = \sum_{i=0}^{n-1} \load[\node_i].
$ 
$\load[\node_i]$ is determined by the quality of the link between $\node_{i}$ and $\node_{i+1}$, as well as the amount of bytes that $\node_{i+1}$ has overheard from $(\node_0, \node_1, ..., \node_{i-1})$. The more it has overheard, the less $\node_{i}$ has to send. Let the erasure ratio and error  ratio of link $\node_{i} \rightarrow \node_{j}$ be $\lera_{i,j}$ and $\lerb_{i,j}$, respectively. With the RS code, the receiving ratio is 
$
\rcvb_{i,j} = (1 - \lera_{i,j}) (1-2\lerb_{i,j}).
$
We maintain the expected number of bytes that are overheard so far at each node, denoted it as $\ovhr[\node_i]$ for node $\node_i$, which is initially 0. Assuming transmissions are independent, $\load[\node_i]$ can be computed iteratively, starting from $\node_0$, shown in Algorithm 1. 
\begin{algorithm}
  \caption{Path Metric Calculation}
  \label{Alg:metric}
  \begin{algorithmic}[1]
	\STATE Set $\ovhr[\node_i] \leftarrow 0$ and $\load[\node_i] \leftarrow 0$ for all $0\le i \le n$. 
	\FOR{$i = 0$ to $n-1$} 
	\STATE $\load[\node_i] \leftarrow \frac{1-\ovhr[\node_{i+1}]}{\rcvb_{i,i+1} \rcvb_{i+1,i}}$ \label{algline:calload} 
	\FOR{$j = i+2$ to $n$} 
	\IF {$\ovhr[\node_{j}] + \load[\node_i] \rcvb_{i,j} \rcvb_{j,i} > 1$} \label{algline:invalid} 
	\RETURN INVALID
	\ENDIF
	\STATE $\ovhr[\node_{j}] \leftarrow \ovhr[\node_{j}] + \load[\node_i] \rcvb_{i,j}$
	\ENDFOR
	\ENDFOR
	\RETURN $\sum_{i=0}^{n-1} \load[\node_i]$	
  \end{algorithmic}
\end{algorithm}
Note that line \ref{algline:calload} calculates the expected number of bytes $\node_i$ should send to $\node_{i+1}$, where $1-\ovhr[\node_{i+1}]$ is the number of bytes still missing at $\node_{i+1}$, and $\rcvb_{i,i+1} \rcvb_{i+1,i}$ is the expected number of bytes that $\node_i$ has to transmit to {\em make sure} that $\node_{i+1}$ receives one byte. The reverse link quality $\rcvb_{i+1,i}$ is considered because $\node_i$ will be expecting the ACK from $\node_{i+1}$ and will transmit again if the ACK is lost. The check in line \ref{algline:invalid} makes sure that the path is valid, because if the condition is true, the path does not have to visit $\node_{i+1}$; otherwise,  $\ovhr[\node_{j}]$ is increased by an amount of   $\load[\node_i] \rcvb_{i,j}$ due to overhearing. With two levels of loops, Algorithm 1's complexity is $O(n^2)$.

\subsection{Routing Algorithm} 
We adopt a greedy algorithm described in Algorithm \ref{Alg:routing} which is similar to the Dijkstra's algorithm. Same as Dijkstra, a set $\djset$ is maintained which keeps the nodes whose paths to the source node have been determined. In each iteration, a node not in $\djset$ is selected that has the shortest path to the source node visiting only nodes in $\djset$. Different from Dijkstra, each node has up to $\djcand$ candidate paths. In each iteration, the candidate paths will be updated when a new node is added to $\djset$. The algorithm returns the best candidate path for each node when terminates. The source node is denoted as  $\node_0$ and the $k_{th}$ candidate path from $\node_0$ to $\node_i$ is denoted as  $\path_k({\node_i})$ where $0 \le k < \djcand$.  The complexity is $O(N^2w)$ where $N$ is the number of nodes in the network.
\begin{algorithm}
  \caption{A Greedy Routing Algorithm}
  \label{Alg:routing}
  \begin{algorithmic}[1]
	\STATE $\djset \leftarrow \emptyset$.
	\STATE $\pmtr[\path_0({\node_i})] \leftarrow \frac{1}{\rcvb_{0,i} \rcvb_{i,0},}$ for all $\node_i$
	\STATE $\pmtr[\path_k({\node_i})] \leftarrow \infty$ for all $\node_i$ and $k$ where $k \neq 0$.
	\WHILE{there are nodes not in $\djset$} 
	\STATE Let $\node_u$ be the node not in $\djset$ with the best candidate path.
	\STATE $\djset \leftarrow \djset \cup \node_u$
	\FOR{all $\node_j$ not in $\djset$} 
	\FOR{$k=0$ to $\djcand-1$}
	\STATE $P \leftarrow \path_k({\node_u}) || \node_j$. 
	\STATE Let $\path_t({\node_j})$ be the candidate path with largest metric. Replace $\path_t({\node_j})$ with $P$ if $\pmtr[\path_t({\node_j})] > \pmtr[P]$.
	\ENDFOR
	\ENDFOR
	\ENDWHILE
  \end{algorithmic}
\end{algorithm}

\section{Error Estimation} 
\label{ErrorEst}

A key component of Crelay is error estimation. A node must know the number of error bytes in a partial packet to be able to ask for the correct number of parity bytes from its upstream node. Our error estimator is referred to as {\em AMPS}, because it is based on the idea of {\em Amplified Sampling}. A naive sampling method would be taking samples of the data bytes and use the ratio of erroneous sampled bytes as an estimate of the error ratio. However, because the raw byte error ratio is usually within [0, 0.2] and is often very small, e.g., 0.01, the naive method may result in high estimation error as it may never sample any erroneous byte. AMPS computes a {\em sample} with multiple bytes which, in effect, amplifies the raw byte error ratio into a much larger sample error ratio, and achieves better estimation accuracy. 


\subsection{The Estimation Procedure}

\begin{table}[t]
\begin{centering}
\begin{tabular}{|c|c|}
\hline 
$\numByte$  & number data bytes in a packet\\
\hline 
$\numB$  & number of segments in a packet\\
\hline 
$\numK$  & number of selected bytes for a sample \\
\hline 
$\numSamp$ & number of samples \\
\hline 
$\rvMisM$  & number of error samples\\
\hline 
$\rvErrB$  & number of error bytes in the packet\\
\hline 
$\rvMaxE$  & maximum number of error bytes in a segment\\
\hline 
\end{tabular}
\par\end{centering}
\caption{List of Notations for AMPS}
\vspace{-0.3in}
\label{table:notation} 
\end{table}

We first assume a frame contains only one data packet with $\numByte$ bytes into $\numB$ segments of equal size. The sender randomly samples $\numK$  data bytes, allowing repeat, and computes their parity bit. Each parity bit is a sample. Clearly, the probability that the sample's parity is flipped is much larger than the probability that a byte has errors. For example, if data byte error ratio is 0.01 and $\numK=32$, the probability that the sample's parity is flipped is $(1-0.99^{32})/2=0.14$, assuming the values of the errors are random. A total of $\numSamp$ samples are calculated in this manner, and the samples are transmitted in the Crelay frame header, protected by error correction code. When the receiver receives the frame, it calculates samples in exactly the same way based on the received data bytes. As some bytes may have been corrupted, the samples it calculates may be different from the samples in the frame header. We call a mismatching sample an {\em error sample}, and denote the number of error samples as $\rvMisM$. $\rvMisM$ carries information about the error conditions in the frame and is used by AMPS as input to calculate the estimate. From a high level, AMPS first finds the maximum a posteriori (MAP) estimation of $\rvErrB$, the number of error data bytes in the packet. It then finds an estimation of $\rvMaxE$, the maximum number of errors in a segment among all transmitted segments, such that the probability that $\rvMaxE$ is greater than its estimate is below a threshold. Table 1 lists the main notations related to AMPS.

The estimation involves three main steps.

\noindent {\em Step 1. The conditional probability $P( \rvMisM = \valMisM | \rvErrB = \valErrB )$.} Note that after interleaving, the error bytes are randomly distributed in the packet. Also, we assume the error bytes take random values. Therefore, the probability that a sample is an error sample when there are $\valErrB$ error bytes, denoted as $\psme_\valErrB$, is 
\begin{equation}
\psme_\valErrB = [1-\left( \begin{array}{c} \numByte-\valErrB \\ \numK  \end{array} \right)/\left( \begin{array}{c} \numByte \\ \numK  \end{array} \right)] / 2 \nonumber
\label{eqpsme}
\end{equation}
For simplicity, we treat the samples as independent. In this case, the probability that there are $\valMisM$ error samples follows the binomial distribution:
\begin{equation}
P(\rvMisM = \valMisM | \rvErrB = \valErrB ) = \left( \begin{array}{c} \numSamp \\ \valMisM \end{array} \right) {\psme_\valErrB}^{\valMisM} (1-\psme_\valErrB)^{\numSamp - \valMisM}  \nonumber
\end{equation}

\noindent {\em Step 2. The MAP estimation of $\rvErrB$.}
$P( \rvErrB = \valErrB  | \rvMisM = \valMisM)$ can be calculated according to the Bayesian formula: 
\begin{equation}
P( \rvErrB = \valErrB  | \rvMisM = \valMisM) = \frac{P(\rvMisM = \valMisM | \rvErrB = \valErrB ) P(\rvErrB = \valErrB) } {\sum_{\valErrB' = 0}^{\numByte} P(\rvMisM = \valMisM | \rvErrB = \valErrB' ) P(\rvErrB = \valErrB')}, \nonumber
\end{equation}
and the MAP estimation $\hat \valErrB$ is the one that maximizes $P( \rvErrB = \valErrB  | \rvMisM = \valMisM)$. Note that this requires the prior distribution of $P(\rvErrB = \valErrB)$, which will be discussed shortly.

\noindent {\em Step 3. The conditional probability $P( \rvMaxE = \valMaxE | \rvErrB = \valErrB)$.} It can be calculated iteratively on the number of segments. To be more specific, we use $P_i( \rvMaxE = \valMaxE | \rvErrB = \valErrB)$ to denote the probability that $\rvMaxE = \valMaxE$ when there are $i$ segments. By definition, $P( \rvMaxE = \valMaxE | \rvErrB = \valErrB) = P_{\numB}( \rvMaxE = \valMaxE | \rvErrB = \valErrB)$. First, when there is only one segment, clearly, 
\[
P_1( \rvMaxE = \valMaxE | \rvErrB = \valErrB) = \left \{ \begin{array} {ll}
   1 & \mbox{$\valMaxE = \valErrB$}\\
   0 & \mbox{ otherwise}
\end{array}     
            \right. 
\]
For notational simplicity, we use $\delta^{\valErrB, i}_{t}$ to denote the probability that among the $\valErrB$ error bytes, $t$ bytes are in one particular segment when there are $i$ segments. Because the error bytes are randomly distributed, 
\begin{equation}
\delta^{\valErrB, i}_{t} = \left( \begin{array}{c} \valErrB \\ t \end{array} \right) (\frac{1}{i})^{t} (1-\frac{1}{i})^{\valErrB - t}  \nonumber
\label{eqblockerr}
\end{equation}
Given $P_i( \rvMaxE = \valMaxE | \rvErrB = \valErrB)$, $P_{i+1}( \rvMaxE = \valMaxE | \rvErrB = \valErrB)$ can be found by conditioning on the number of error bytes in a tagged segment. That is, we single out one segment, called the tagged segment, and check the number of error bytes in this segment. The event that $\rvMaxE = \valMaxE$ occurs when (1) the tagged segment has less than $\valMaxE$ errors while the maximum number of errors in a segment among the remaining segment is exactly $\valMaxE$, or (2) the tagged segment has exactly $\valMaxE$ errors while the maximum number of errors in a segment among the remaining segments is no more than $\valMaxE$. Therefore,
\begin{eqnarray}
\vspace{-0.05in}
P_{i+1}(\rvMaxE = \valMaxE | \rvErrB = \valErrB) &=& \sum_{t=0}^{\valMaxE-1} \delta^{\valErrB, i}_{t} P_{i}(\rvMaxE = \valMaxE | \rvErrB = \valErrB - t) \nonumber \\
&+& \delta^{\valErrB, i}_{\valMaxE} \sum_{\valMaxE' = 0}^{\valMaxE} P_{i}(\rvMaxE = \valMaxE' | \rvErrB = \valErrB - \valMaxE) \nonumber
\vspace{-0.05in}
\end{eqnarray}

After getting $P( \rvMaxE = \valMaxE | \rvErrB = \valErrB)$ as well as $\hat \valErrB$, AMPS outputs $\hat \valMaxE$  such that $P( \rvMaxE \le \hat \valMaxE| \rvErrB = \hat \valErrB)$ is greater than a threshold, set to be 0.95 in our implementation. Finally, if there are multiple data packets in the frame, AMPS first estimates the total number of errors in the whole frame and divides it into each packet proportional to the packet sizes, then estimates of maximum number of errors in a segment for each packet.

\subsection{AMPS In Practice}

\subsubsection{Multiple Resolutions}
For the estimation of $\rvErrB$, we introduce AMPS with multiple resolutions, because a particular $\numK$ will always be best for one range of error ratios, but may over-amplify or under-amplify others. We use three types of samples with $\numK = 128$, $\numK = 32$, and $\numK = 10$, referred to as the type-0, type-1, and type-2 samples, respectively. The type-0, type-1, and type-2 samples are responsible for data error ratios in the range of $[0.001, 0.01]$, $[0.01, 0.03]$, and $[0.03, 0.2]$, respectively. Assuming the error values are random, they amplify the data error ratios roughly into sample error ratios of $[0.06, 0.36]$, $[0.14, 0.31]$, and $[0.13, 0.45]$, respectively, such that every range is sufficiently amplified to allow enough number of error samples to be drawn. The number of samples for type-0, type-1, and type-2 samples are 8, 16, and 40, respectively, which are determined by considering the required number of outputs in each range. The number of errors in a typical packet with error ratio range of $[0.001, 0.01]$, $[0.01, 0.03]$, and $[0.03, 0.2]$ are below 20 bytes, 60 bytes, and 400 bytes, respectively, and being able to output respectively 9, 17, and 41 different values should suffice. We basically let each estimator run in parallel. In cases when the three estimators give different estimates, we take a conservative approach and pick the largest estimate. Note that the total number of samples is 64, or only 8 bytes.

\subsubsection{Table Lookup Implementation}
To reduce the computation complexity, AMPS obtains the estimation with a constant time, simple table lookup after $\rvMisM$ is computed. Note that $P( \rvMaxE = \valMaxE | \rvErrB = \valErrB)$ can be precomputed such that the estimation of $\rvMaxE$ can be found by a table lookup given the MAP estimation of $\rvErrB$. Because $P( \rvMisM = \valMisM | \rvErrB = \valErrB )$ can be precomputed, the MAP estimation of $\rvErrB$ can also be obtained by a table lookup given the value of $\rvMisM$, if the distribution of $\rvErrB$ is fixed. However, in practice, the distribution of $\rvErrB$  can vary depending on the wireless channel condition. We cope with it by computing tables for 100 representative distributions. As the size of the frame and the number of segments may vary, we also choose representative sizes of the frame and segment and compute corresponding tables. For any received frame under any channel condition, the table that is closest to its parameters is chosen. All such tables are computed for each value of $\numK$. The total size of the tables used by AMPS is about 1.5MB in the current implementation and can be further reduced after relaxing the accuracy requirements.

\subsubsection{Acquiring $P(\rvErrB = \valErrB)$} 


AMPS requires the prior $P(\rvErrB = \valErrB)$, which should be estimated and selected among the representative distributions. We tested the Cisco Aironet 802.11a/b/g wireless cardbus adapter \cite{ciscocard} and found that, interestingly, the byte error ratio distribution can be fitted very well in many cases by the truncated Pareto distribution: 
\[
P(x) = \frac{\alpha \gamma^{\alpha} x^{-\alpha-1}}{1-(\gamma/\nu)^{\alpha}},
\]
where $\gamma$ and $\nu$ are the lower limit and the upper limit of $x$, respectively, and $\alpha$ controls the heaviness of the tail \cite{jasa06_Pareto}. This may be due to the heavy tailed nature of the error burst in the received packets. Therefore, the distribution can be described by only one parameter $\alpha$, as $\gamma$ and $\nu$ are known in practice, i.e., $\gamma$ is a very small number and $\nu$ is a number close to 1. As 100 distributions are needed, we precompute tables for $\alpha$ from 0.02 to 2 at a step of 0.02, while $\gamma=0.001$, $\nu=0.999$.

\comment{
\begin{figure}[h]
\begin{center}
\includegraphics[width=2.5in]{54.eps} 
\end{center}
\caption{The distributions of byte error numbers in 1500-byte packets transmitted as 54Mbps. The $Y$-axis is cut off at 0.01 to show the details of the curve.}
\label{fig:byteerror}
\vspace{-0.2in}
\end{figure} 
}
To select a distribution, $\alpha$ should be estimated. We implemented the estimation method in \cite{jasa06_Pareto}, where we record the error ratios of the received frames, denoted as $\lerb_i$ for frame $i$. Suppose there are $n$ recent samples. According to Equation (4) in \cite{jasa06_Pareto}, the estimation, denoted as $\tilde \alpha$, should satisfy
\[
\frac{n}{\tilde \alpha} + \frac{n (\gamma/\nu)^{\tilde \alpha} \ln{\gamma/\nu} }{1-(\gamma/\nu)^{\tilde \alpha}} + n \ln{\gamma}  = \sum_{i=1}^{n} \ln{\lerb_i}.
\]
Given the set of error ratio values, we find $\tilde \alpha$ that results in a left side of the above equation closest to the measured $\sum_{i=1}^{n} \ln{\lerb_i}$ on the right side of the equation. 

\subsection{Comparing with EEC}
\label{compEEC}

The other error ratio estimator we are aware of is EEC \cite{EEC}, which also uses the parity bit of multiple bits as a sample and uses multiple levels of samples. We independently arrived at a similar idea used in AMPS. Besides that, the approaches of AMPS and EEC profoundly differ. AMPS is designed according to the standard procedure in estimation theory, namely the MAP estimation, which is optimal for minimizing the average estimation error required by our application. EEC is based on a simple algorithm without using the knowledge of the prior distribution of error ratios. We note that the prior distribution can usually be measured at the receiver. Both AMPS and EEC have low computational overhead. However, for 1500-byte packets, the overhead of AMPS and EEC are 8 bytes and 36 bytes, respectively. 

\begin{figure}
\begin{center}
\vspace{-0.1in}
\includegraphics[width=2.3in]{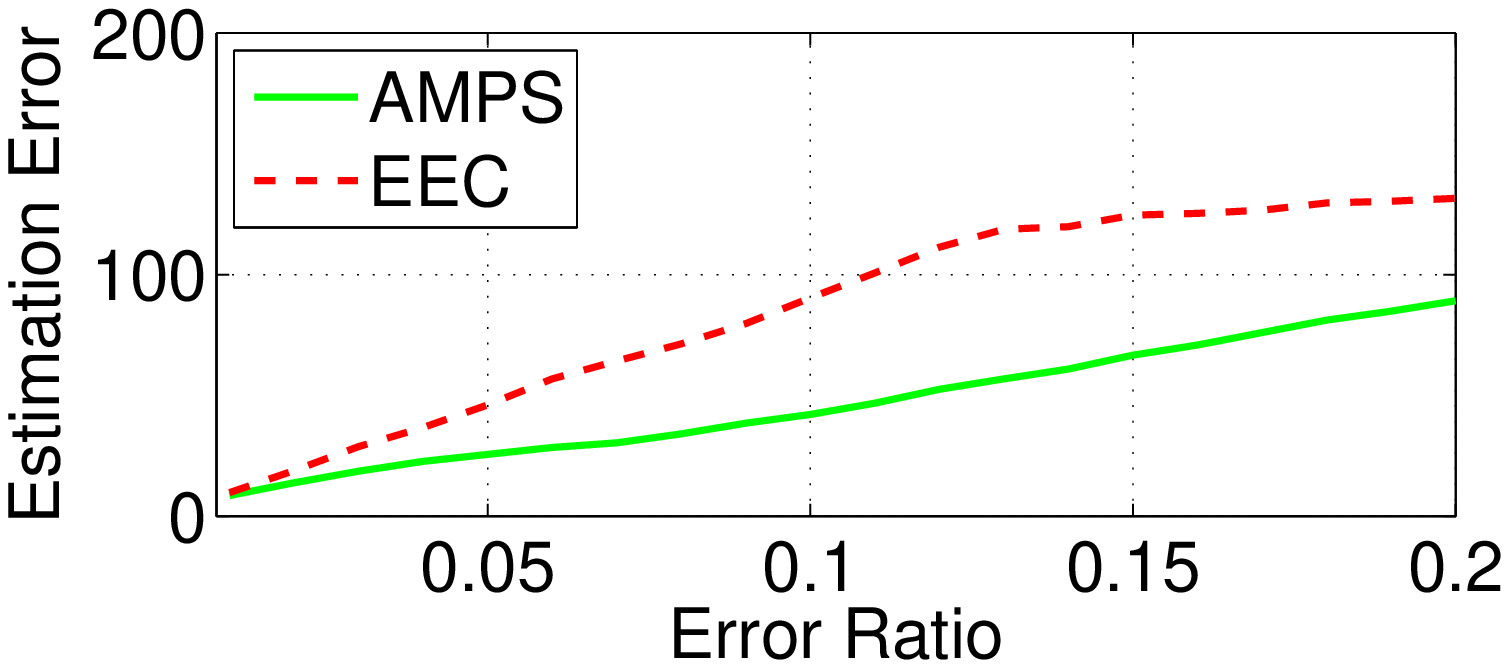}\\ 
\end{center}
\vspace{-0.2in}
\caption{Comparison of estimation error between AMPS and EEC.}
\label{fig:AMPSEEC}
\vspace{-0.2in}
\end{figure}

To compare the per-packet level estimation performance, we implemented EEC and ran simulations. As AMPS and EEC estimate byte and bit errors, respectively, we injected random byte and bit errors into 1500-byte and 1500-bit packet for them where the error ratio ranged from 0.01 to 0.2 at a step of 0.01. For each ratio the simulation ran for 10,000 times. As EEC assumes no prior knowledge, for a fair comparison, AMPS was also not given the prior, and ran based on the distribution of error ratio with the heaviest tail in which case AMPS approximates a maximum likelihood estimator. The performance is measured by the difference between the estimated and the real number of injected error bytes/bits. Fig.~\ref{fig:AMPSEEC} suggests the that AMPS outperforms EEC significantly. We believe the reason is that EEC may discard certain samples while such samples still carry much information. On the other hand, without the prior distribution, the maximum likelihood estimator is the optimal estimator. We show in Section \ref{Evaluation} the performance of AMPS in real-life experiments when the prior knowledge is available.

\section{Experimental Results}
\label{Evaluation}

We conducted experiments to evaluate the performance of Crelay. The compared schemes include 
\begin{itemize}
\item {\em More}: the benchmark opportunistic routing protocol,
\item {\em Srcr}: the benchmark traditional routing protocol, where nodes use the shortest path according to the ETX metric to forward packet hop-by-hop without exploiting partial packets and overhearing.
\end{itemize}
We used the original implementation at \cite{MORESite} for More and used our own implementation for Srcr.

\subsection{Implementation}

We implemented a prototype of the Crelay protocol in around 5,000 lines of C++ code within the Click modular router \cite{Click} as a user space daemon. Packets are transmitted in broadcast frames, same as More \cite{MORE}. The main addition to the Crelay protocol discussed in Section \ref{protocol} is a simple mechanism to cope with interference. Basically, we allow a node $\node_A$ to send a polling message to another node $\node_B$, if $\node_A$ has not heard any message from $\node_B$ for longer than a threshold, while $\node_A$  has many packets in state $S_1$ for $\node_B$ or has many packets in state $S_2$ for $\node_B$ that have been transmitted but have not been ACKed. Once heard the polling message, $\node_B$ transmits while others backoff for a time. The reason is that $\node_B$'s transmission could be lost in collisions due to hidden terminal problems. This approach is adopted because RTS/CTS is not designed for broadcast packets in 802.11. 

We also made two optimizations. First, we allow nodes to send parity bytes preemptively, i.e., parity bytes are sent along with data bytes in the first transmission attempt, if the link has non-zero error ratio. As a result, many packets with errors can be decoded at the receiver without requiring the overhead of feedback. In our current implementation, if the error ratio is $\lerb > 0$, a node transmits bytes that can correct $\max\{0.02,\min\{0.05,\lerb\}\}$ fractions of errors for the bytes it sends.  Second, in AMPS, we set the minimum number of estimated errors in a segment to be 3, as this does not increase much overhead but can reduce the underestimation probability.

\subsection{Testbed and Experiment Setup}
\begin{figure}
\begin{center}
\includegraphics[width=1.2in]{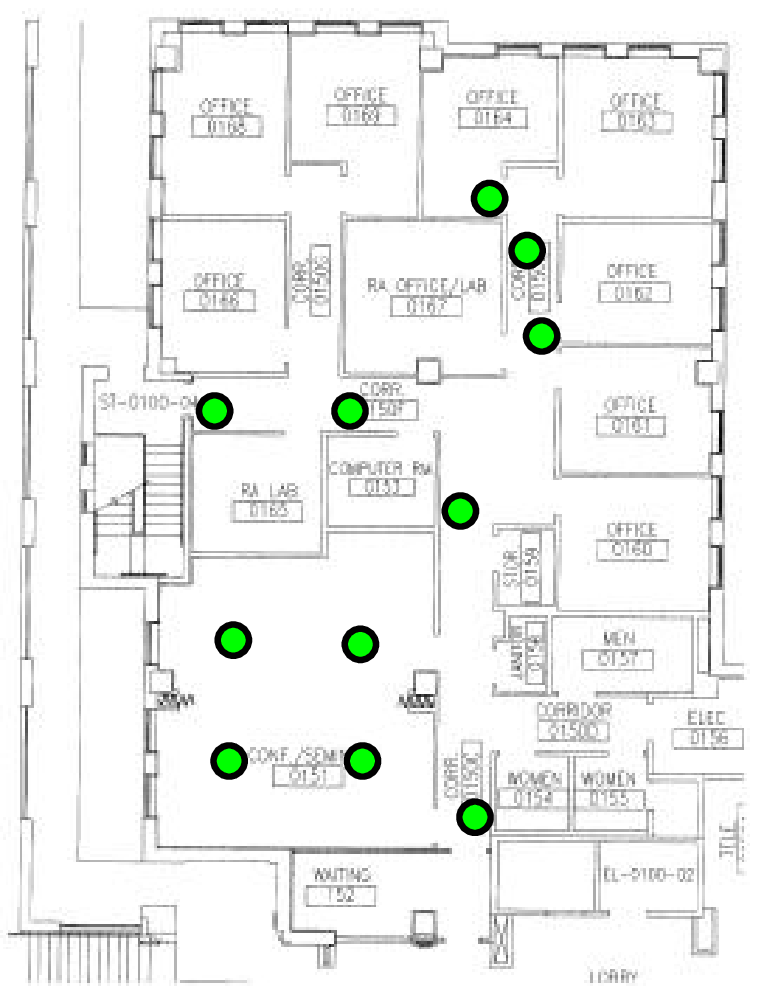} \hspace{0.3in}
\includegraphics[width=1.2in]{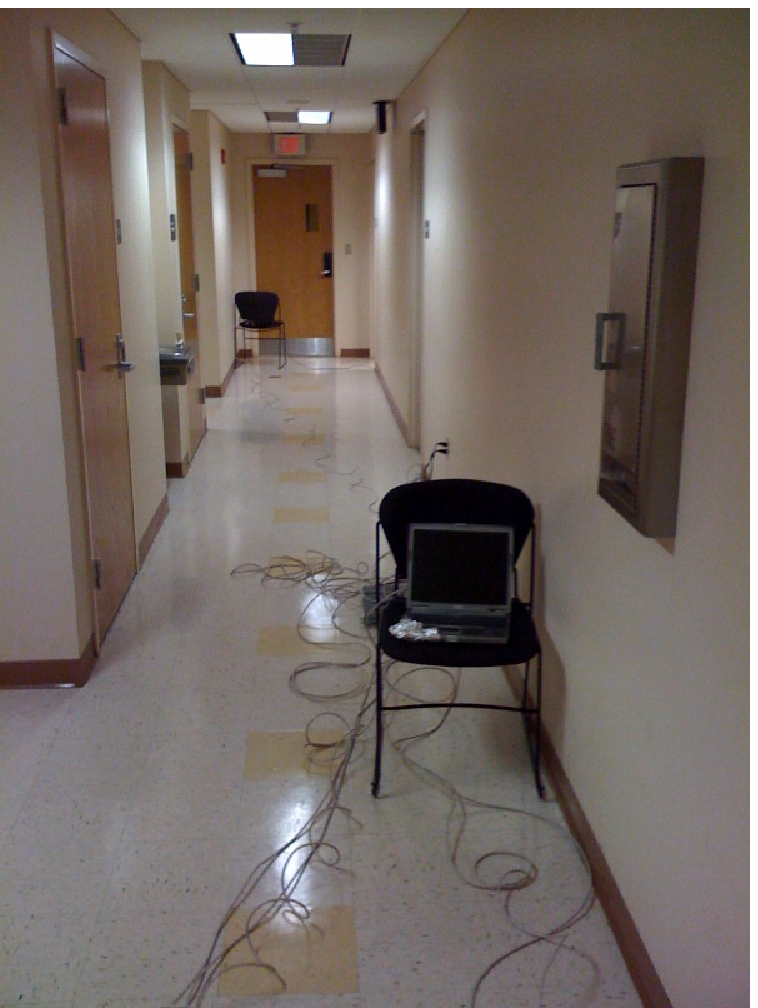}
\end{center}
\vspace{-0.15in}
\caption{The setup of the testbed.}
\label{fig:testbed}
\vspace{-0.15in}
\end{figure}

We employed an 11-node wireless testbed. The wireless nodes are laptop computers with Cisco Aironet 802.11a/b/g cardbus adapter. In our experiments, the wireless cards ran on 802.11b/g channel 3 (2.427GHz) when there was little traffic (less than 3 beacon packets/sec) during the experiments. The transmission power was set to be 1dBm; in addition, aluminum foil was wrapped around the card to reduce the transmission/reception power to allow the experiments to be carried out within the confinements. The testbed setup, as well as two of machines used in the experiments, are shown in Fig.~\ref{fig:testbed}. With this set up, we were able to get 46 links where we consider a link exists if the erasure ratios of both directions are lower than 0.8. Among the links, the average RSSI was 6.98dB, the average erasure ratio was 0.271, and the average error ratio was 2\%. We ran the Madwifi \cite{Madwifi} driver in the monitor mode, which allows us to receive raw data frames for exploiting partial packets. The MTU was set to be 2200 bytes and the data rate 1Mbps. 

We selected 72 pairs of nodes in the network, between which Crelay found the path to be more than one hop. This selection was made because we are interested in the performance of multi-hop paths. Each pair is a flow, and we ran the experiment when only one flow was active. In each experiment, nodes first learn the link states by sending Hello packets in the first 9 seconds at random intervals, where a node sends around 50 packets in total. In the next 9 seconds, similar to More \cite{MORESite}, the link state is propagated with the help of a central node using wired links. At time 18 second, the topology learning phase ends and nodes begin to send data packets. The source node generates packets of size 1500 bytes every 10ms and the experiment runs for another 10 seconds. 

\subsection{Performance}

\begin{figure}
\begin{center}
\includegraphics[width=2.3in]{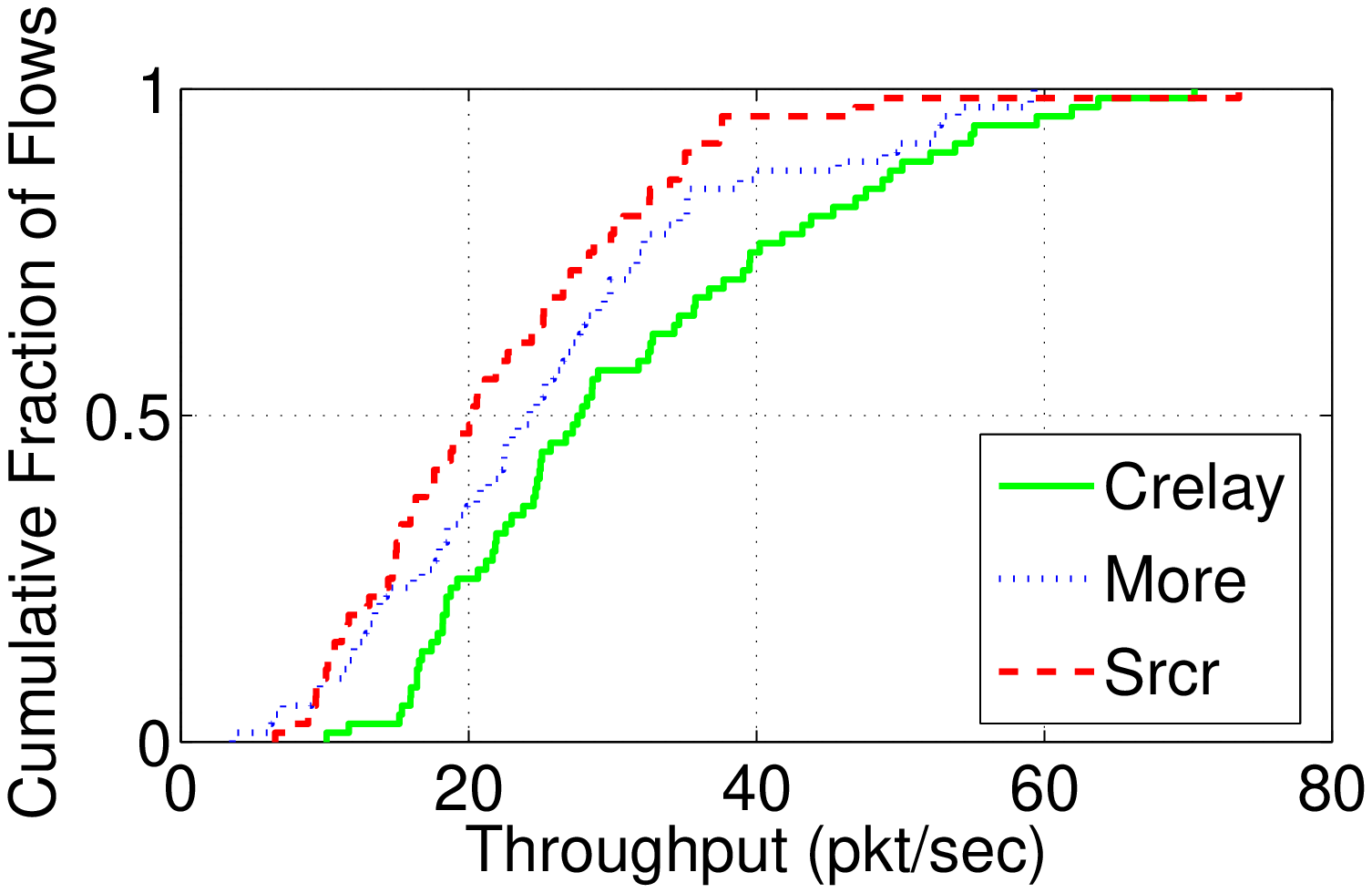}\\ 
\end{center}
\vspace{-0.2in}
\caption{The cumulative distribution of flow throughput.}
\label{fig:throughput}
\vspace{-0.2in}
\end{figure}

We first show in Fig.~\ref{fig:throughput} the cumulative distribution of the throughput of Crelay, More, and Srcr measured in the number of received packets by the destination  per second. We can see that Crelay has a significantly higher throughput than both More and Srcr. Fig.~\ref{fig:singlef} reveals more interesting details, which shows the scattered plots of Crelay v.s. More and Crelay v.s. Srcr for each flow. In the scattered plot, a point the 45-degree line represents a flow in which two compared schemes have the same throughput. Fig.~\ref{fig:singlef} shows that Crelay outperforms More in most flows and outperforms Srcr in almost all the flows. As a quantitative measure, for each flow, we define the throughput gain of scheme A over scheme B as $(\mu_A-\mu_B) / \mu_B$, where $\mu_A$ and $\mu_B$ are the throughput of scheme A and scheme B, respectively. We found that average throughput gain of Crelay over More is 36\% and the average throughput gain of Crelay over Srcr is 52\%.

\comment{
\begin{figure}
\begin{center}
\includegraphics[width=2in]{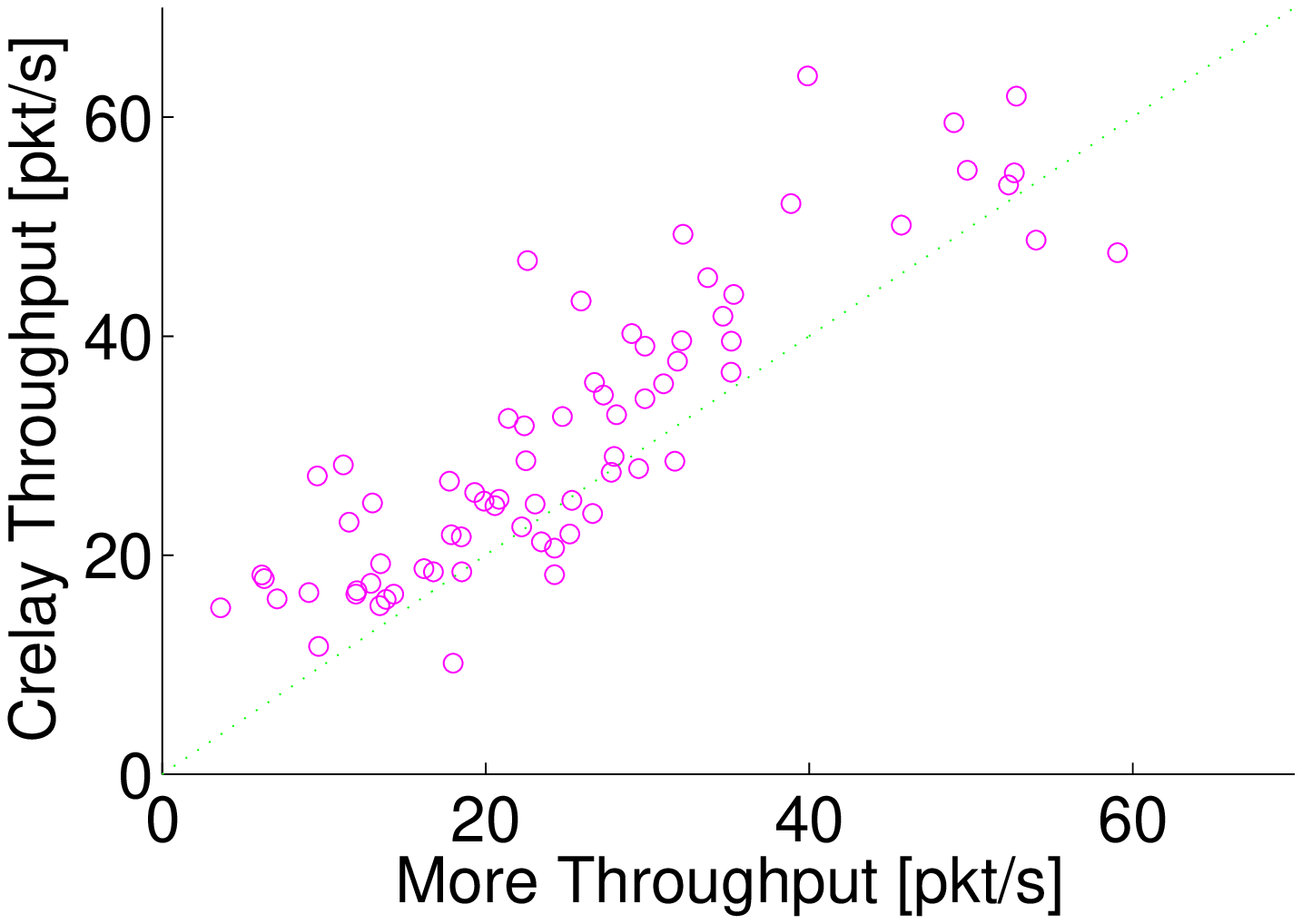}\\ 
\vspace{-0.05in}
{\small (a)} \\
\vspace{-0.01in}
\includegraphics[width=2in]{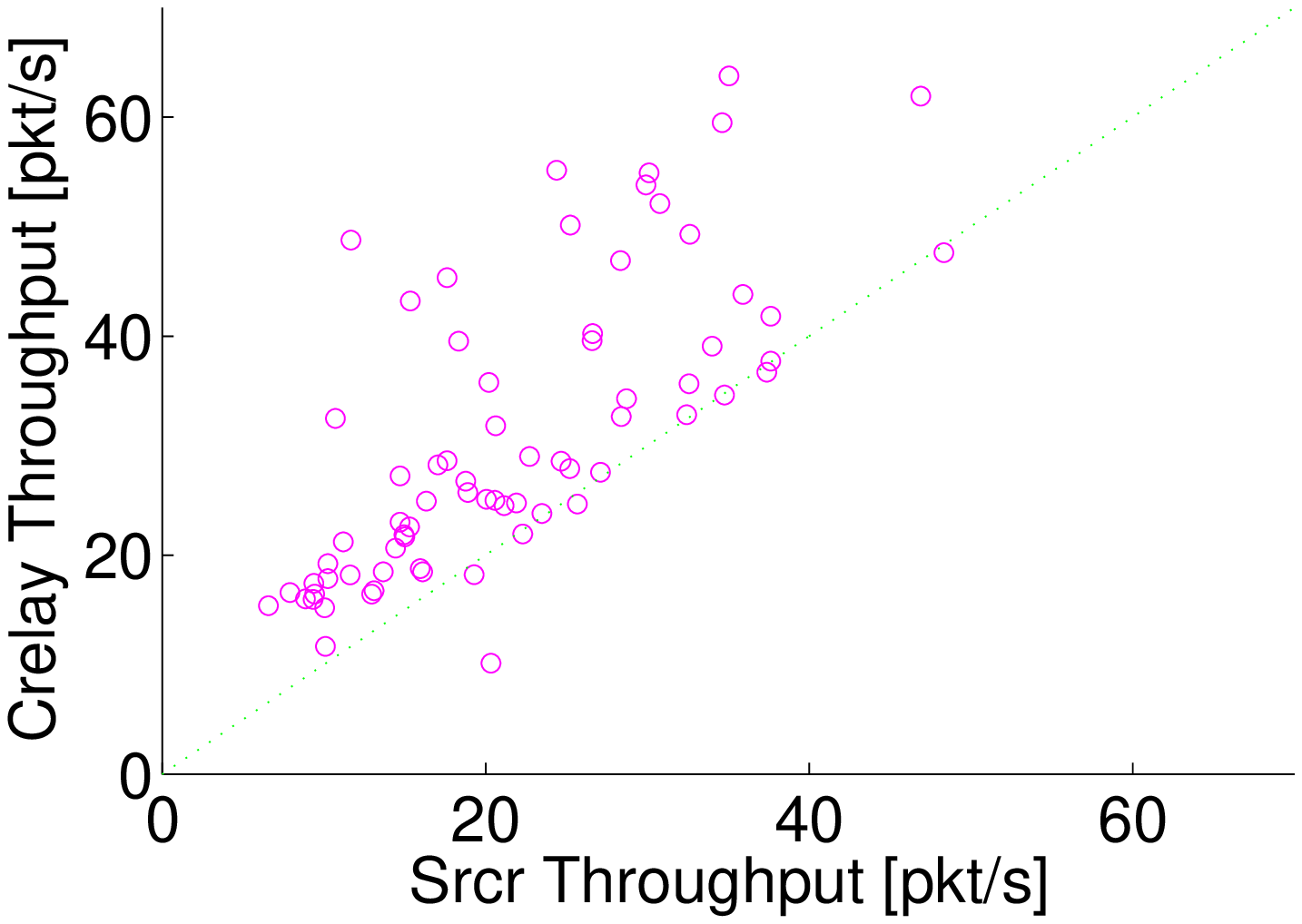}\\ 
\vspace{-0.05in}
{\small (b)}
\vspace{-0.03in}
\end{center}
\vspace{-0.1in}
\caption{The scattered plot comparison of the flow throughput.}
\label{fig:singlef}
\vspace{-0.2in}
\end{figure} 
}

\begin{figure}
\begin{center}
\vspace{0in}
\includegraphics[width=1.7in]{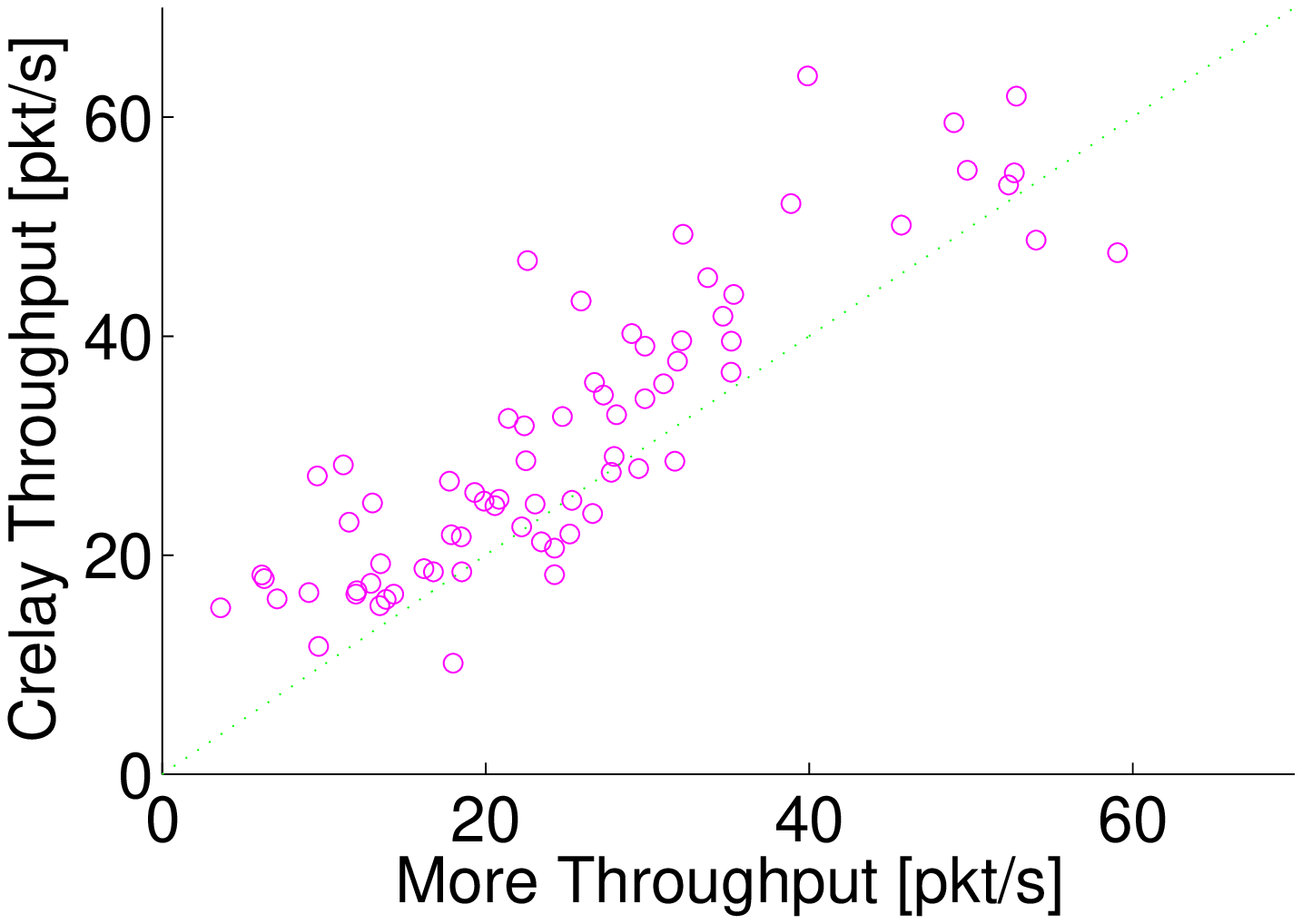} \hspace{0.01in}
\includegraphics[width=1.7in]{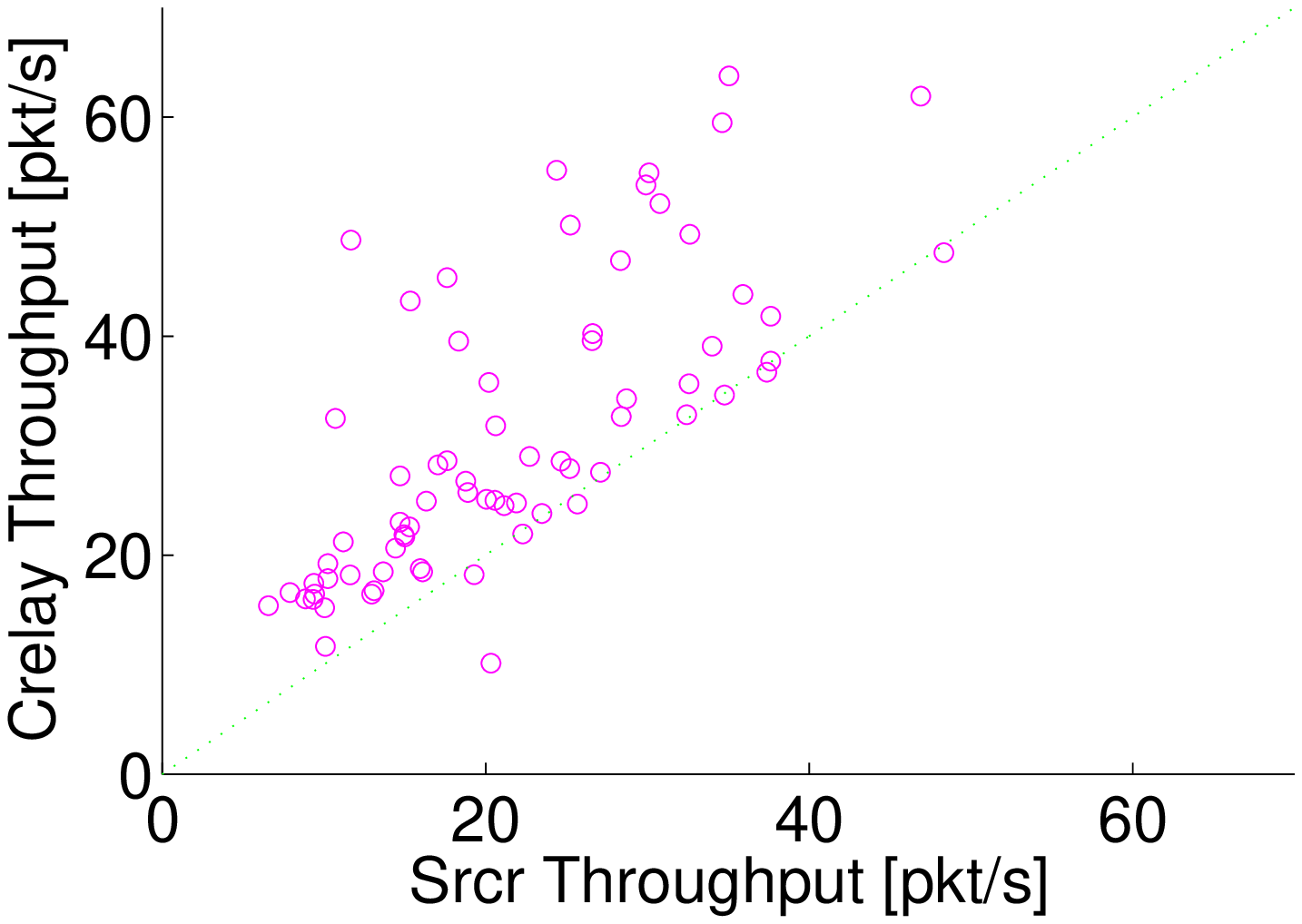}\\ 
\vspace{-0.01in}
\end{center}
\vspace{-0.15in}
\caption{The scattered plot comparison of the flow throughput.}
\label{fig:singlef}
\vspace{-0.15in}
\end{figure}

\begin{figure}[h]
\begin{center}
\vspace{-0.15in}
\includegraphics[width=2.3in]{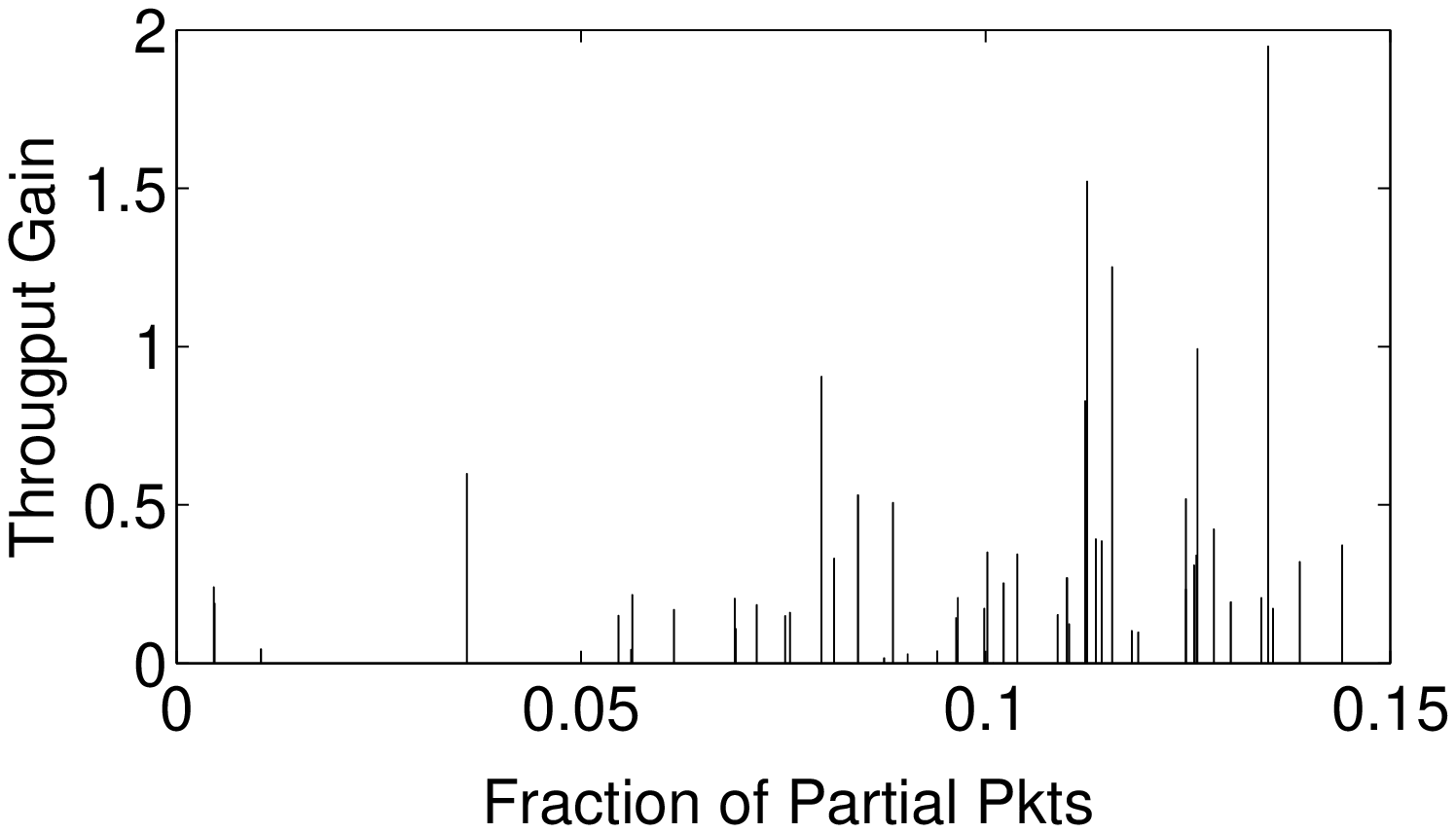}\\ 
\end{center}
\vspace{-0.23in}
\caption{The flow throughput gain and the fraction of partial packets.}
\label{fig:ppct}
\vspace{-0.2in}
\end{figure}

One of the gains of Crelay is from exploiting partial packets. Fig.~\ref{fig:ppct} shows the relation of gain and partial packet ratio for each flow, where the $x$ axis is the percentage of partial packets received by the nodes on the packet forwarding path when running Crelay, and $y$ axis is the throughput gain of Crelay over More. We can see that there is a positive correlation between the gain and the percentage of the partial packets.

\begin{figure}[h]
\begin{center}
\vspace{-0.1in}
\includegraphics[width=2.3in]{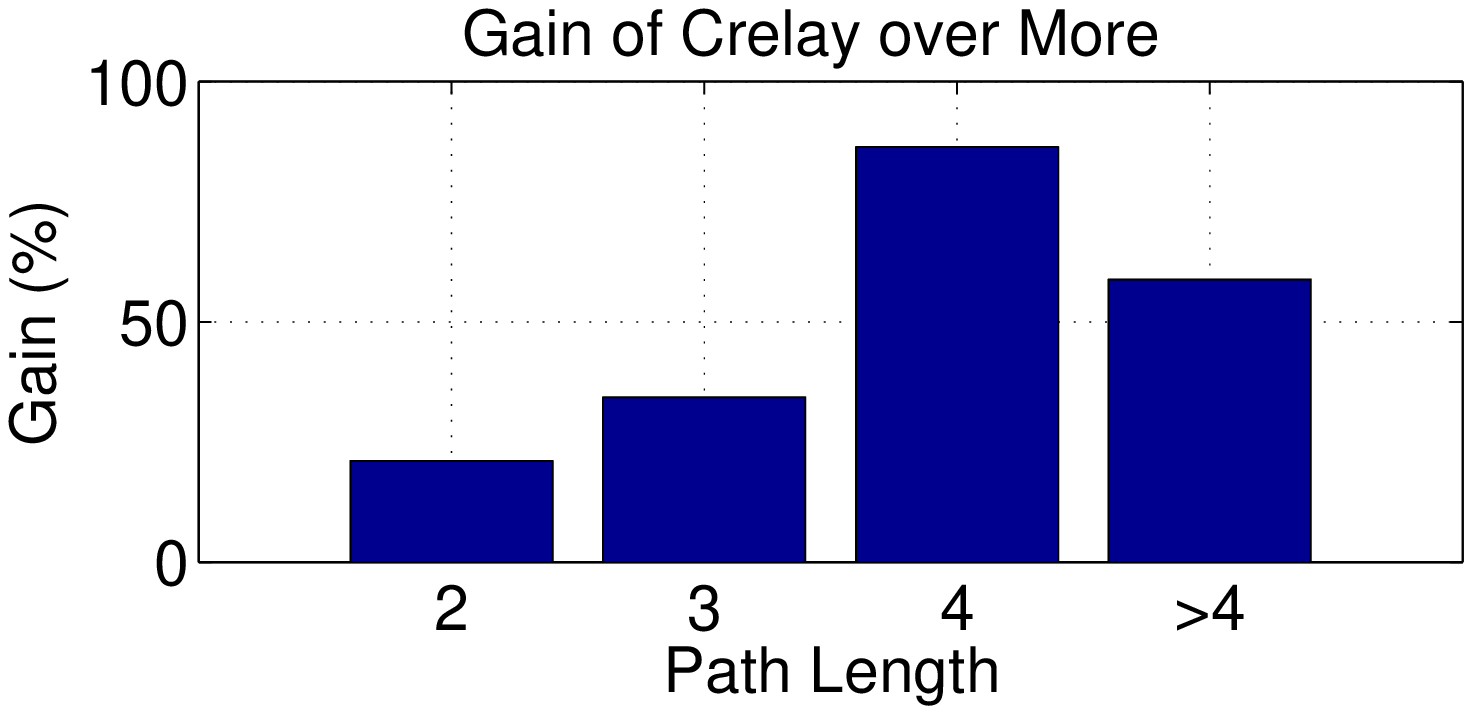}\\ 
\end{center}
\vspace{-0.23in}
\caption{Throughput gain for different path lengths.}
\label{fig:gainplen}
\vspace{-0.1in}
\end{figure}

Fig.~\ref{fig:gainplen} shows the average throughput gain of Crelay over More for paths of different lengths, where the path length is based on the path used by Crelay. The number of paths are 34, 26, 10 and 2 for path lengths of 2, 3, 4, and greater than 4, respectively. We can see that the gain is usually higher for longer paths, which may be because such paths usually have weaker channels and more opportunities to exploit partial packets. 

\begin{figure}[h]
\begin{center}
\vspace{-0.2in}
\includegraphics[width=2.3in]{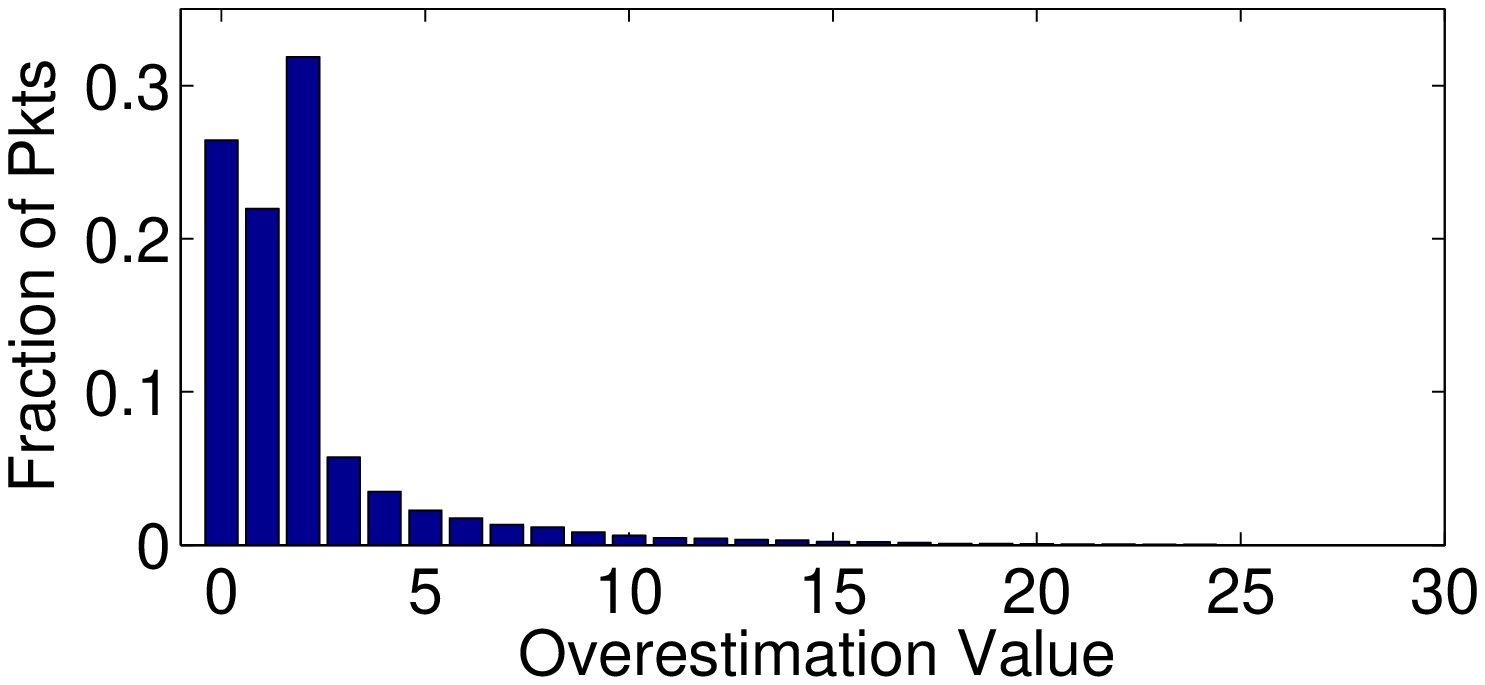}\\ 
\end{center}
\vspace{-0.23in}
\caption{The p.d.f. of overestimation of AMPS.}
\label{fig:AMPSOverE}
\vspace{-0.1in}
\end{figure}

The performance of Crelay is largely dependent on AMPS. With our current choices of parameters, we found that for 23.7\% of the times, AMPS underestimates the number of errors and more parity bytes have to be transmitted. However, even in such cases, usually most of the blocks are decoded and only a few blocks need more transmissions, because the numbers of errors in the records are different. For 3.89\% of the time, AMPS underestimates the number of errors, but the available parity bytes, sent preemptively, are actually sufficient for correcting all the errors. For the rest of the cases AMPS overestimates, and Fig.\ref{fig:AMPSOverE} shows the probability density function of the number of overestimation. We can see that if overestimated, for more than 80.1\% of the times, AMPS overestimates by no more than 3 bytes per codeword. It is possible to tune the parameters to achieve other underestimation/overestimation tradeoff.

\comment{ 
We also study the packet delay with Crelay. It may be a concern that Crelay may increase the packet delay as an upstream node may hold a packet for transmission until the receiving status of the packet at the next hop is known. However, when the network is loaded, the main delay is due to queuing. To verify this, we show the cumulative distribution of the packet delay of Crelay and Srcr in Fig.~\ref{fig:delay}. The packet delay of More is not shown because it is not included in its experiment report. We can see that there is virtually no difference between the delay of Crelay and Srcr. 
\begin{figure}
\begin{center}
\includegraphics[width=2.3in]{delaycmp.eps}\\ 
\end{center}
\caption{The cumulative distribution of delay of Crelay and Srcr.}
\label{fig:delay}
\vspace{-0.2in}
\end{figure} 
}

\section{Conclusion}
\label{Conclusions}

In this paper, we proposed Coded Relay (Crelay) for multi-hop wireless networks. With Crelay, nodes can exploit partial packets and overhearing for packet forwarding. One feature of Crelay is that nodes can often send some parity bytes to the next hop to recover the packet, which is significantly smaller than the size of the packet. We proposed and implemented the Crelay protocol in software. We studied the routing problem with Crelay and proposed a greedy algorithm for finding paths. We also designed an error ratio estimator, called AMPS, that can estimate the number of errors in a packet with good accuracy at very low overhead. We tested Crelay on an 11-node testbed, and the results show that Crelay is capable of achieving significant gain over existing protocols.

\end{document}